\documentclass{aa}  

\usepackage{graphicx}
\usepackage{txfonts}
\usepackage{multirow}
\usepackage{color}
\bibpunct{(}{)}{;}{a}{}{,}

\newcommand{\kms}{km\,s$^{-1}$}
\newcommand{\vsini}{$v\sin i$}
\newcommand{\vsinia}{$v\sin i_{\rm A}$}
\newcommand{\vsinib}{$v\sin i_{\rm B}$}
\newcommand{\vrad}{$v_{\rm rad}$}
\newcommand{\vmic}{$v_{\rm mic}$}
\newcommand{\vmac}{$v_{\rm th}$}
\newcommand{\msun}{M$_{\odot}$}

\newcommand{\bl}{$B_{\ell}$}
\newcommand{\bd}{$B_{\rm D}$}
\newcommand{\ddeg}{$^{\circ}$}
\newcommand{\te}{$T_{\rm eff}$}
\newcommand{\logg}{$\log{g}$}
\newcommand{\rr}{$R_A/R_B$}
\newcommand{\lr}{$L_A/L_B$}
\newcommand{\prot}{$P_{\rm rot}$}

\begin{document}

   \title{The magnetic field of the double-lined spectroscopic binary system HD 5550\thanks{Based on the BinaMIcS Large Program (PI: C. Neiner, runID: L131N02) obtained at the Telescope Bernard Lyot (USR5026) operated by the Observatoire Midi-Pyr\'en\'ees, Universit\'e de Toulouse (Paul Sabatier), Centre National de la Recherche Scientifique of France.}}
   \titlerunning{The magnetic field of HD 5550 in the framework of the BinaMIcS project}

   \author{E. Alecian\inst{1,2,3}
          \and
          A. Tkachenko\inst{4}
          \and
          C. Neiner\inst{3}
          \and
          C.P. Folsom\inst{1,2}
          \and
          B. Leroy\inst{3}
          \and
          the BinaMIcS collaboration
          }

   \institute{Univ. Grenoble Alpes, IPAG, F-38000 Grenoble, France
         \and
         CNRS, IPAG, F-38000 Grenoble, France
         \and
         LESIA, Observatoire de Paris, PSL Research University, CNRS, Sorbonne Universit\'es, UPMC Univ. Paris 06, Univ. Paris Diderot, Sorbonne Paris Cit\'e, 5 place Jules Janssen, 92195 Meudon, France
         \and
         Instituut voor Sterrenkunde, KU Leuven, Celestijnenlaan 200D, 3001 Leuven, Belgium
             }

   \date{Received September 15, 1996; accepted March 16, 1997}

 
  \abstract
   {{ The origin of fossil fields in intermediate- and high-mass stars is poorly understood, as is the interplay between binarity and magnetism during stellar evolution.  Thus we have begun a study of the magnetic properties of a sample of intermediate-mass and massive short-period binary systems, as a function of binarity properties. }}
   {{ This paper specifically aims to characterise the magnetic field of HD 5550, a double-lined spectroscopic binary system of intermediate-mass.}}
   {We gathered 25 high-resolution spectropolarimetric observations of HD 5550 using the instrument Narval. We first fit the intensity spectra using Zeeman/ATLAS9 LTE synthetic spectra to estimate the effective temperatures, microturbulent velocities, and the abundances of some elements of both components, as well as the light-ratio of the system. We then applied the multi-line least-square deconvolution (LSD) technique to the intensity and circularly polarised spectra, which provided us with mean LSD $I$ and $V$ line profiles. We fit the Stokes $I$ line profiles to determine the radial and projected rotational velocities of both stars. We then analysed the shape and evolution of the $V$ profiles using the oblique rotator model to characterise the magnetic fields of both stars.}
   {We confirm the Ap nature of the primary, previously reported in the literature, and find that the secondary displays spectral characteristics typical of an Am star. While a magnetic field is clearly detected in the lines of the primary, no magnetic field is detected in the secondary, in any of our observation. If a dipolar field were present at the surface of the Am star, its polar strength must be below 40~G. The faint variability observed in the Stokes $V$ profiles of the Ap star allowed us to propose a rotation period of $6.84_{-0.39}^{+0.61}$\,d, close to the orbital period ($\sim$6.82\,d), suggesting that the star is synchronised with its orbit. By fitting the variability of the $V$ profiles, we propose that the Ap component hosts a dipolar field inclined with the rotation axis at an angle $\beta=156\pm17$\,\ddeg\ and a polar strength $B_{\rm d}=65 \pm 20$ G. The field strength is the weakest known for an Ap star.
   }
   {}


   \maketitle
%

\section{Introduction}

The last decades brought a wealth of information on the magnetic fields of A-, B- and O-type stars. It is now well established that a small fraction (5-10\%) of OBA stars host strong ($\sim$100~G to $\sim$30~kG) fossil (i.e. not continuously maintained against ohmic decay) fields organised on large-scales (dominated by low-order multipolar fields). These fields are observed in stars of all masses from $\sim$2~\msun\ to $\sim$40\msun\, and all evolutionary stages from the pre-main sequence to the end of the main sequence, and even maybe in the red-giant phase \citep{donati09,alecian13a,auriere15,wade15}. The favoured hypothesis assumes that these fields are fossil and have been shaped during the star formation \citep{Borra82,moss01}. While this hypothesis can explain almost all the properties of the OBA magnetic fields, we still need to understand why only a small fraction of the OBA stars host fossil fields, while the large majority host very weak or no fields.

The impact of the magnetic fields on the structure and environment of OBA stars can be of different order depending on the magnetic strength and the temperature of the stars. On one hand, in A-type stars the magnetic fields seem to affect only the composition of the upper layers, producing strong chemical  stratification and surface peculiarities. This stratification is due to the magnetic field impact on diffusion processes as observed in the magnetic Ap/Bp stars \citep{michaud70,babel91,aleciang10,stift12}. On the other hand, magnetic fields in O and early-B type stars have a strong impact on the close stellar environment by forming complex and highly dynamical magnetospheres within a few stellar radii, and could even modify significantly the mixing processes in their interiors \citep{uddoula02,uddoula08,townsend05,briquet12}. Fossil fields can also affect significantly the angular momentum evolution of the stars. In early B- and O-type stars the magnetic coupling between the radiatively-driven winds and the magnetic fields can carry away a large amount of angular momentum, braking the surface of the stars, and forcing an internal redistribution of angular momentum \citep{uddoula09}. Observational evidence of rotational braking has been recently obtained in one magnetic B star \citep[$\sigma$~Ori~E,][]{townsend10}. In late-B and A-type stars, a strong dichotomy exists between the non-magnetic single stars that rotate more rapidly than the magnetic ones \citep{abt95}. Whether such a dichotomy has been shaped during the early pre-main sequence phase or even earlier remains to be understood \citep{stepien00,alecian13b}.

The presence of a companion, in some cases, can also have an effect on the evolution of stars. An initially eccentric binary system with non-synchronised, non-aligned components will tend to an asymptotic state with circular orbits, synchronised components, and aligned spins \citep[e.g.][]{hut80}. When both objects are originally close enough, such an evolution can occur on time-scales shorter than the star lifetime, and produces slowly rotating stars \citep{zahn77}. Furthermore such an evolution is possible thanks to a dissipation of the kinetic energy of internal flows into heat. Whether or not such flows affect the fossil fields residing in OBA stars is still an open question \citep{ogilvie07,remus12,cebron14}. In addition, in the case of companions separated by only a few stellar radii, and depending on the mass of the stars, interactions such as magnetospheric interaction, wind-wind interaction, magnetic reconnection, or magnetic braking - orbit interaction can occur, affecting the environnement and evolution of the stars \citep{russell11,uddoula09,uddoula14,gregory14,barker09}. Finally, it is now evident that magnetic fields play a major role on the star formation, and in particular on the core fragmentation, hence the formation of binary and multiple systems \citep{commercon11,masson12}. Theoretical studies in recent years have shown that magnetic flux must be removed from the original molecular cloud to be able to reproduce the commonly observed binary and multiple systems, and young stars with disks. When and how this removal occurs, and whether it is at the origin of the magnetic dichotomy of the OBA stars remains to be understood.

It is in this context that the Binarity and Magnetic Interaction in various classes of Stars (BinaMIcS\footnote{http://binamics.lesia.obspm.fr/}) project { has taken place,} in order to study the interplay between magnetism and binarity, and their impact on the structure, environment and evolution of the stars. { To this aim}, we need to gather information on the basic statistics and the magnetic field strength and topology of a statistically large sample of close binary systems { (with orbital periods shorter than 20 days).  These are the systems in which we expect significant mutual interaction via tidal or magnetospheric interaction. 

The system studied in this paper is \object{HD 5550}. It} was previously reported to be an Ap SrCrEu type star \citep{renson91}. Based on a measurement of the Geneva Z-index \citep{cramer80}, \citet{north84} give a photometric estimation of the mean surface field of about 100~G, i.e. one of the faintest fields of their sample. Its SB2 nature was discovered by \citet{carrier02} thanks to CORAVEL and ELODIE data for a sample of chemically peculiar stars obtained to study multiplicity among CP stars. The analysis of their data allowed them to constrain the orbit. Assuming a null eccentricity, they find an orbital period around 6.8 days, and a mass ratio of about 1.5. They also report chemical peculiarities in both components of the system, but cannot distinguish between Am and Ap peculiarities.

In this paper we report on our analysis of the intensity and polarised spectra of HD 5550. We first describe the observations in Section 2. Then in Section 3 we detail our analysis of the intensity spectra that allowed us to estimate the effective temperatures and chemical peculiarities of both stars, as well as the orbital parameters. The analysis of the polarised spectra, and how it allowed us to constrain the surface magnetic fields of both components is detailed in Section 4. Finally, a summary is given in Section 6.

\begin{table}
\caption{Log of the observations.}
\label{tab:log}      
\centering          
\begin{tabular}{rcccr}
\hline\hline       
Sp. \# & Date          & HJD             & $t_{\rm exp}$  & S/N \\
          & (y-m-d) UT & (2450000+)  & (s)                   &         \\
\hline
 1 & 2013-07-12 01:31  & 6485.58380 & 3600 & 780 \\
 2 & 2013-07-15 01:45  & 6488.59387 & 3600 & 890 \\
 3 & 2013-08-05 01:00  & 6509.56377 & 3600 & 820 \\
 4 & 2013-08-09 02:02  & 6513.60659 & 3600 & 860 \\
 5 & 2013-08-10 02:00  & 6514.60584 & 3600 & 900 \\
 6 & 2013-08-11 02:36  & 6515.63015 & 3600 & 910 \\
 7 & 2013-08-12 02:09  & 6516.61210 & 3600 & 910 \\
 8 & 2013-08-14 02:37  & 6518.63129 & 3600 & 920 \\
 9 & 2013-08-15 03:13  & 6519.65647 & 3600 & 950 \\
10 & 2013-08-16 02:12  & 6520.61433 & 3600 & 950 \\
11 & 2013-08-18 02:53  & 6522.64248 & 3600 & 890 \\
12 & 2013-08-19 02:57  & 6523.64561 & 3600 & 840 \\
13 & 2013-08-20 01:49  & 6524.59818 & 3600 & 950 \\
14 & 2013-08-21 02:46  & 6525.63789 & 3600 & 800 \\
15 & 2013-08-22 02:58  & 6526.64664 & 3600 & 930 \\
16 & 2013-09-02 23:51  & 6538.51762 & 3600 & 810 \\
17 & 2013-09-05 03:23  & 6540.66447 & 3600 & 870 \\
18 & 2013-09-09 02:58  & 6544.64753 & 3600 & 900 \\
19 & 2013-09-11 02:39  & 6546.63425 & 3600 & 910 \\
20 & 2013-09-12 01:18  & 6547.57834 & 3600 & 930 \\
21 & 2013-09-14 00:50  & 6549.55916 & 3600 & 680 \\
22 & 2013-09-16 01:18  & 6551.57837 & 3600 & 930 \\
23 & 2013-09-18 00:49  & 6553.55839 & 3600 & 780 \\
24 & 2013-09-20 01:50  & 6555.60065 & 3600 & 950 \\
25 & 2013-09-21 00:41  & 6556.55317 & 3600 & 960 \\
\hline                  
\end{tabular}
\end{table}

\section{Observations}

Twenty five observations have been obtained with the instrument Narval installed at the T\'elescope Bernard Lyot (TBL, Pic du Midi, France), in the circular polarimetric mode, as part of a two-year Large Program allocated to the BinaMIcS collaboration (PI: C. Neiner) and started in March 2013. Each polarimetric measurement was obtained by acquiring four successive individual spectra, between which we rotated the retarders in an appropriate way so as to exchange the beam pass of both orthogonal polarisation states in the 2nd and 3rd observations. This allows to minimise the systematic errors due to small instrumental defects. All measurements have been reduced using the dedicated data reduction software Libre-ESpRIT \citep{donati97}, available at the telescope. The resulting spectrum covers a wavelength range from 370 to 1048 nm, with a resolution of 65000 \citep[for additional information on the Narval instruments and data, see e.g.][]{donati06,donati08,silvester12}.

The intensity $I$ component of the Stokes vector is formed by adding the four individual spectra, while the polarised Stokes $V$ component has been obtained by combining them using the ratio method \citep{bagnulo09}. A diagnostic null spectrum $N$ has also been obtained by combining the spectra in such a way as to cancel the polarisation from the object, which may reveal possible spurious polarisation contributions \citep{donati97}. Finally, a continuum normalisation of the Stokes $I$ spectra has been performed order by order to all observations using the IRAF\footnote{IRAF is distributed by the National Optical Astronomy Observatories, which are operated by the Association of Universities for Research in Astronomy, Inc., under cooperative agreement with the National Science Foundation} routine \emph{continuum} by fitting cubic spline functions to the line-free portions of the spectra. Table 1 summarises the log of the observations. Columns 1 to 5 give the spectrum identification, date and UT time, heliocentric Julian date (HJD) of the middle of the observation, total exposure time, and signal-to-noise ratio (S/N) of the observations per spectral pixel at 597\,nm in the $V$ spectrum.


\section{Analysis of the intensity spectra}

\subsection{Spectral disentangling}

We use the Fourier-based spectral disentangling ({\sc spd},
hereafter) technique \citep{hadrava95} as implemented in the {\sc
FDBinary} code \citep{ilijic04} to simultaneously optimise the
disentangled spectra of both stellar components and the orbital
elements of the system. The Fourier implementation of the method is
superior to the original technique introduced by \citet{simon94}
applied in the wavelength domain. It is indeed much faster
and is thus readily applicable to the (long) time-series of
high-resolution spectroscopic data.

\subsubsection{Orbital solution}

\begin{figure}[t]
\includegraphics[scale=0.94]{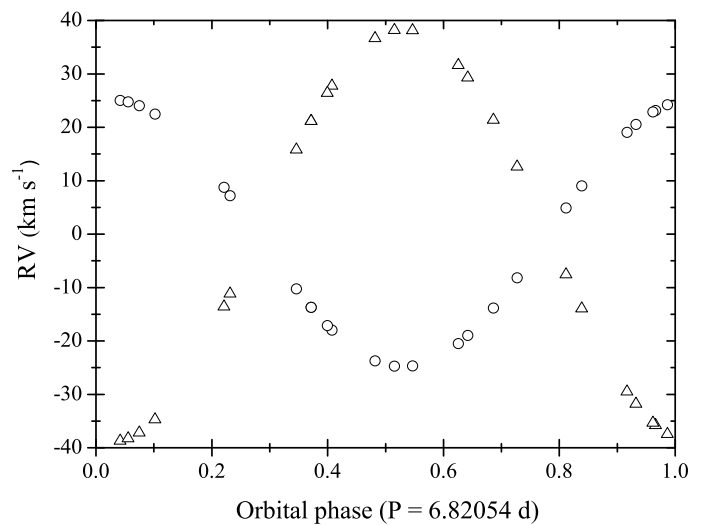}
\caption{Orbital phase distribution of the spectra of HD\,5550 in
terms of the RVs computed from our final orbital solution (see
Table~\ref{Table1}). The primary and secondary RVs are shown by
circles and triangles, respectively. Phase zero corresponds to the
maximum RV separation of the two stars.} \label{Figure1}
\end{figure}

Since the {\sc spd} method becomes degenerate when the orbital
period is optimised together with other orbital elements, we have
chosen to fix the period to the value of 6.82054~d reported by
\citet{carrier02}. Our choice is justified given the precision of
2$\times10^{-4}$~d on the orbital period that has been reached in
the above mentioned study. For the {\sc spd} solution to be stable
and unique, a uniform spectral coverage of the orbital period is
required. As illustrated in Fig.~\ref{Figure1}, our spectra provide
excellent phase coverage, with only a small gap between orbital phases
0.1 and 0.2, with phase zero being indicative of the maximum radial
velocity (RV, hereafter) separation of the two stars.

We selected six wavelength regions between 4000 and 5000 \AA\ from our composite spectra and
used them to optimise the solution for the following five orbital
elements: the time and longitude of periastron (T$_0$ and $\omega$),
the RV semi-amplitudes of both stellar components (K$_1$ and K$_2$),
and the orbital eccentricity ($e$). The selection of the wavelength
regions used for the calculation of the orbital solution was made
very carefully to ensure they contain a sufficient number of strong
metal lines that can be cut at the local continuum points. The orbital elements obtained from the individual wavelength
regions as well as their average values are reported in
Table~\ref{Table1}. Our orbital solution is in good agreement with
the solution presented by \citet{carrier02}. We follow Carrier et al. (2002) and define the primary (A) star as the hottest and more massive one, and the secondary (B) star as the coolest and less massive one.

\begin{figure*}
\centering
\includegraphics[scale=1.1]{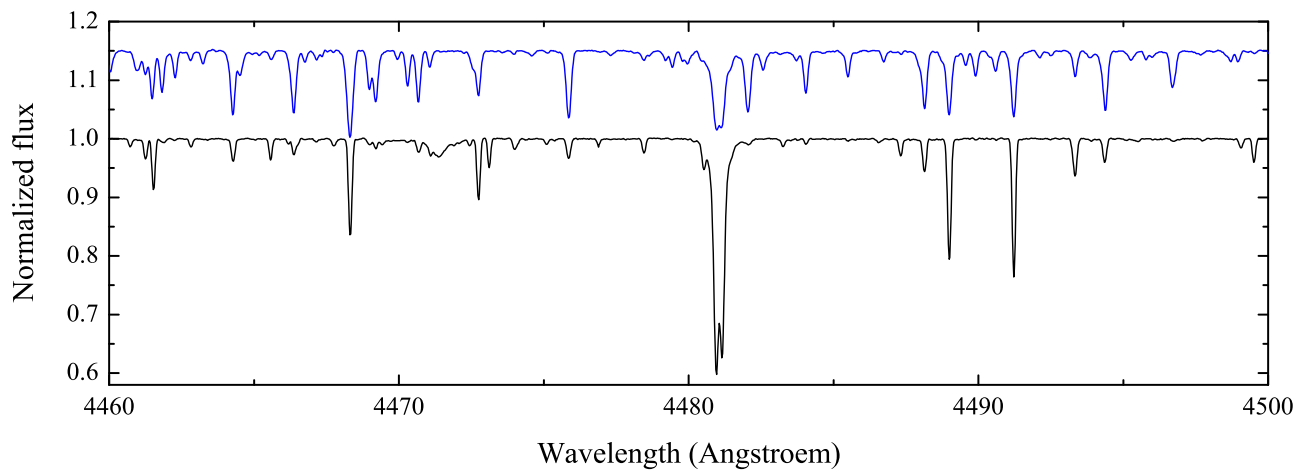}
\caption{A small part of the disentangled spectra of the components of
HD\,5550. The spectrum of the secondary (top, in blue) was shifted compared to the spectrum of the primary (bottom, in black) for better visibility.} \label{Figure2}
\end{figure*}

\begin{table*}
\centering\tabcolsep 2.0mm\caption{Orbital elements of the HD\,5550
system computed from six individual wavelength regions. Our final
orbital solution is represented by the average values computed from
these six regions; errors are 1$\sigma$ standard deviations of the
mean and are given in parenthesis in terms of last
digits.}\label{Table1}
\begin{tabular}{lllllllll} \hline
\multirow{2}{*}{Parameter} & \multirow{2}{*}{Unit} & \multicolumn{6}{c}{Wavelength range (\AA)\rule{0pt}{11pt}} & \multicolumn{1}{c}{\multirow{2}{*}{Average}}\\
& & 4042--4047 & 4125--4133 & 4170--4175 & 4557--4560 & 4822--4825 & 4916--4929\\
\hline
$P$\rule{0pt}{11pt} & days & 6.82054$^*$ & 6.82054$^*$ & 6.82054$^*$ & 6.82054$^*$ & 6.82054$^*$ & 6.82054$^*$ & 6.82054(20)$^*$\\
T$_0$ & HJD (24\,000\,00+) & 50988.42 & 50988.50 & 50988.47 & 50988.43 & 50988.53 & 50988.43 & 50988.46(5)\\
$e$ & & 0.005 & 0.005 & 0.006 & 0.007 & 0.008 & 0.006 & 0.006(1)\\
$\omega$ & degree & 167.1 & 171.1 & 169.7 & 168.1 & 173.1 & 167.9 & 169.5(2.3)\\
K$_{\rm A}$ & \kms & 24.97 & 25.28 & 24.96 & 24.86 & 24.80 & 24.92 & 24.97(17)\\
K$_{\rm B}$ & \kms & 38.72 & 38.48 & 38.43 & 38.57 & 38.66 & 38.60 & 38.58(11)\\
\hline
\multicolumn{9}{l}{$^*$adopted from \citet{carrier02}. In the last column the uncertainty in bracket is from \cite{carrier02}.\rule{0pt}{11pt}}\\
\end{tabular}
\end{table*}

\subsubsection{Disentangling of stellar spectra}

We use the orbital solution reported in the last column of
Table~\ref{Table1} to compute the disentangled spectra of the
individual stellar components of the HD\,5550 system. The spectra
are disentangled over a wide wavelength range, from 4150 to 5700~\AA, and
excludes the spectral region of the H$_{\beta}$ line. This line is
not considered because of the pronounced interstellar band
present in the red wing of the profile. We do not go further to the
blue part of the spectrum because the uncertainty in the
normalisation of the observed spectra rises towards the blue edge
of the spectra. The red part of the spectrum is omitted because of
large contribution from telluric lines. The spectral disentangling
technique is well-known to suffer from undulations in the resulting
decomposed spectra \citep{hadrava95,ilijic04}, which is due to
the instability of the zero frequency mode in the Fourier domain. To
minimize the effect, we divide our spectra into $\sim$30~\AA\
overlapping wavelength regions, and merge them afterwards to produce
the disentangled spectra in the complete wavelength range. The
remaining low-amplitude undulations are corrected by fitting
low-degree polynomial to a few carefully selected continuum points
in both disentangled spectra before the merging of the individual
wavelength regions. A small portion of the separated spectra of both
components is illustrated in Fig.~\ref{Figure2}.

\subsection{Fundamental parameters determination}


We have fit the combined intensity spectrum of the system to determine many stellar parameters: the radial, microturbulent, and projected rotational velocities (\vrad, \vmic, \vsini), the surface abundances, the effective temperatures (\te), and the light-ratio (\lr). The continuum normalisation of our echelle spectra is too approximate in the wings of the Balmer lines to constrain the surface gravities. We have fixed them to appropriate numbers for main-sequence A-type stars ($\log g=4.0-4.5$).
The fitting procedure we have adopted consists of several steps. We first determined approximate guesses of the parameters by fitting with the eye the combined spectrum of one observation. We then applied a semi-automatic fit to derive more accurate values of a set of parameters: the effective temperatures, microturbulent velocities, abundances, and light-ratio. Next we used these values to compute line-masks appropriate to both components and applied the LSD technique that computes weighted cross-correlated profiles. We then fit the LSD profile shapes to determine accurately the radial and rotational velocities. Finally, we applied once more the semi-automatic fitting procedure using these new values and determined new values of the effective temperatures, microturbulent velocities, abundances, and light-ratio. We find that the effective temperatures have not significantly changed and that the line-masks do not need to be modified, which closes the iteration process. We detail below the successive steps.

\subsubsection{The eye-fitting procedure}

To get initial estimates of the effective temperatures, \vsini, and radial velocities of the two components, we first performed a fit with the eye using the IDL visualisation script BINMAG1 (kindly put at our disposal by O. Kochukhov).
\mbox{BINMAG1} computes the composite spectrum of a binary star, and takes as input two synthetic spectra of different effective temperatures and gravities, each corresponding to one of the two binary components. The code convolves the synthetic spectra with instrumental, turbulent and rotational broadening profiles. It then combines them according to the ratio of radii of the components specified by the user, and the flux ratio given by the atmospheric models, to produce the spectrum of the binary star. The individual synthetic spectra were calculated in the local thermodynamic equilibrium (LTE) approximation, using the code SYNTH3  (Kochukhov 2007). SYNTH3 requires, as input, atmosphere models obtained using the ATLAS~9 code (Kurucz 1993) and a list of spectral line data obtained from the VALD database\footnote{http://ams.astro.univie.ac.at/$\sim$vald/} (Vienna Atomic Line Data base, VALD3 version). In this process we fixed the surface gravity to $\log g=4.0$, and the microturbulent velocities to 2\,\kms, respectively. We modified the parameters that are affecting most the spectra: the effective temperatures, \vsini, radial velocities of both components, and ratio of radii, until a match was approximately found. We first assumed solar abundances, then modified the abundances of some elements showing the strongest anomalies (Cr, Fe, Ti) to reach a better match. We emphasise that this procedure was employed only to estimate first guesses, allowing the semi-automatic procedure described below to work efficiently.


\begin{figure*}[t]
\centering
\includegraphics[width=3.5cm,angle=270,bb = 91 53 222 715,clip]{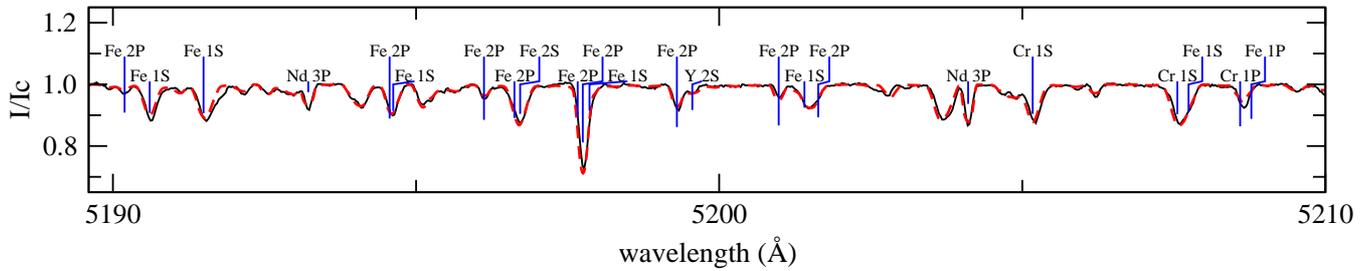}
\caption{Observation \#1 (black solid line) superimposed with a Zeeman synthetic spectrum (dashed red line) computed with the parameters of Table \ref{tab:fitfp} and $\log g = 4.0$. The main transition from the primary (P) or the secondary (S) are indicated with vertical blue bars.}
\label{fig:fitsp}
\end{figure*}

\subsubsection{The semi-automatic fitting procedure}

To improve the fit and get a better estimates of the parameters, we then used the LMA semi-automatic routine developed by one of us (CPF), and described in detail in \citet{folsom12}. LMA performs an iterative fit using a Levenberg-Marquardt $\chi^2$ minimisation technique and composite synthetic spectra of a binary star. The synthetic spectra of both stars are computed using the ZEEMAN spectrum synthesis program \citep{landstreet88,wade01}, given the effective temperatures, surface gravities, rotational and microturbulent velocities for each star. An additional local thermal broadening (\vmac) can also be set to non-zero values. We initially set it to 0~\kms. We assumed negligible effects of the magnetic fields on the spectral line of the primary, which is reasonable considering the relatively faint magnetic polar strength of the primary, and the low upper limit on the magnetic polar strength of the secondary (Sect. 4). Both synthetic spectra are computed in absolute flux. Then they are shifted to the appropriate radial velocities, and they are added, weighted by the square of the ratio of radii specified by the user. The spectra are then normalised by the sum of the continuum spectra of the two stars, weighted by the square of the ratio of radii.


The semi-automatic fitting procedure we have adopted consists in several steps. Assuming at first solar abundances, we have alternatively fitted the parameters of one of the two components (\te, \vmic) and the ratio of radii (\rr) while fixing the parameters of the other component, until a convergence was found (i.e. until a relative difference, between one iteration and the next, below 3\% was found for all the parameters). For both stars, the rotational broadening and radial velocities were fixed to the eye-fitting estimates. For each iteration, we have fitted simultaneously 17 spectral windows of 100\,\AA\ maximum width, distributed between 4160\,\AA\ and 6460\,\AA, and avoiding the Balmer lines and the portions of the spectra heavily contaminated with telluric lines. We then compared the result of the fit with the observation, and computed the local $\chi^2$ value for each spectral bin, as follows:
\begin{equation}
\chi^2_i = \left(\frac{I_{i\,\rm obs}-I_{i\,\rm mod}}{\sigma(I_{i\,\rm obs)}}\right)^2
\end{equation}
This allows one to estimate the significance of the difference between the observation and the model, and identify the elements with the strongest anomalies. We found that the abundances of some elements (see Table \ref{tab:fitfp}) needed to be adjusted for both stars, in order to get a better agreement with the observations. To this aim we proceeded as above by fitting the parameters of each star, alternatively, until a convergence was reached. For each star, we fit the abundances ($\log N_{\rm X}/N_{\rm H}$) first, by fixing the fundamental parameters (\te, \vmic, and \rr), and then fixed the abundances to the new ones, and fit once more the fundamental parameters. We proceeded this way as many time as needed until convergence was reached. 
The whole procedure was applied one more time using  the more accurate values of \vsini, \vrad\ and \vmac, derived from the mean LSD profiles (Sect. 3.5).

\subsubsection{Adopted parameters and uncertainty estimation}

The uncertainties on the derived parameters can have many sources: uncertainties on \logg, \vsini, \vmac\, and spectral variability. Indeed, Ap stars often show rotational variability in their spectra due to inhomogeneous abundance patches over the stellar surface or to rotational modulation of the Zeeman effect in individual lines. We can therefore expect such variability in our spectra, which would imply uncertainties on the fundamental parameters obtained using only one observation at one rotation phase.  Additional uncertainties could be caused by the blends of the primary and secondary lines, which change with the orbital phase. To estimate the maximum effect on the fitted parameters, we selected two observations obtained at different orbital phases, and rotational phases. The rotation period of the magnetic star, being yet unknown, we based our choice on the amplitude of the Zeeman signature of the $V$ observations. We fixed our choice to the spectra \#1 and \#11, and performed many semi-automatic fits by fixing the \logg\ values to 4.0 and 4.5, and the \vsini\ and \vmac\ to their minimum and maximum values given by their error bars (Table \ref{tab:fitfp}). The average, minimum and maximum values of all the solutions have been used to determine the adopted values and uncertainties on the effective temperatures, microturbulent velocities, light-ratio, and abundances, summarised in Table \ref{tab:fitfp}. Our study confirms that the primary component is an Ap star, and we find that the secondary component is an Am star.


\begin{table}[!ht]
\centering
\caption{Adopted fundamental parameters}
\begin{tabular}{l|l|cc}
\hline\hline
\multicolumn{2}{l|}{Component} & A & B \\
\hline
\multicolumn{2}{l|}{\rr} & \multicolumn{2}{c}{$1.24_{-0.23}^{+0.04}$} \\ [5 pt] 
\multicolumn{2}{l|}{\vsini\ (\kms)} & $4.7_{-2.8}^{+1.3}$ & $3.2_{-1.9}^{+0.3}$ \\ [5 pt] 
\multicolumn{2}{l|}{\vmic\  (\kms)} & 0.0(8) & $3.2_{-1.9}^{+0.3}$ \\ [5 pt] 
\multicolumn{2}{l|}{\vmac\  (\kms)} & $1.9_{-1.9}^{+1.5}$ & $4.0_{-2.9}^{+1.3}$ \\ [5 pt] 
\multicolumn{2}{l|}{\te\  (K)} & 11400(300) & $7800_{-80}^{+170}$ \\ [5 pt] 
\multicolumn{2}{l|}{\logg\  (cgs)} & \multicolumn{2}{c}{[4.0,4.5]} \\ 
\hline
\multirow{11}{*}{$\log N_{\rm X}/N_{\rm H}$} & C & & -3.89(19) \\
& Mg & -5.23(15) & \\
& Ca & & -5.996(17) \\
& Sc & & -8.934(24) \\
& Ti & & -6.83(5) \\
& Cr & -5.72(11) & -5.96(6) \\
& Mn & -6.28(18) & \\
& Fe & -4.36(11) & -4.513(9) \\
& Ba & & -8.84(23) \\
& Pr & -8.18(17) & \\
& Nd & -8.01(17) & \\
\hline
\end{tabular}
\label{tab:fitfp}
\tablefoot{The parameters that have been fitted in the fitting procedure are those with the uncertainties indicated in brackets. The uncertainties refer to the last digits of the associated values. The parameters that have been fixed have no uncertainties. The two fixed values of \logg\ used in the procedure are indicated.}
\end{table}

\begin{figure*}[!ht]
\centering
\includegraphics[width=6cm]{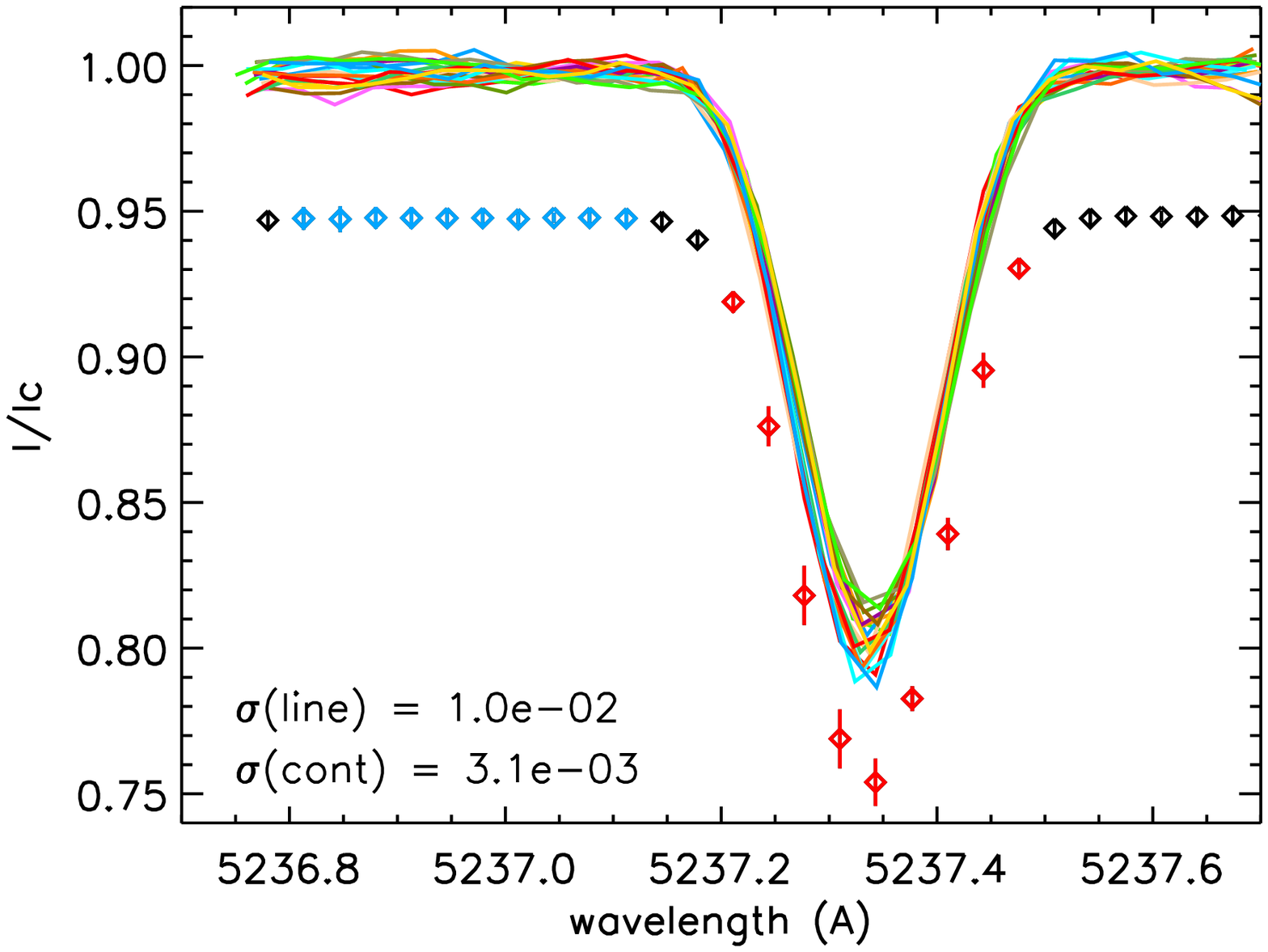}
\includegraphics[width=6cm]{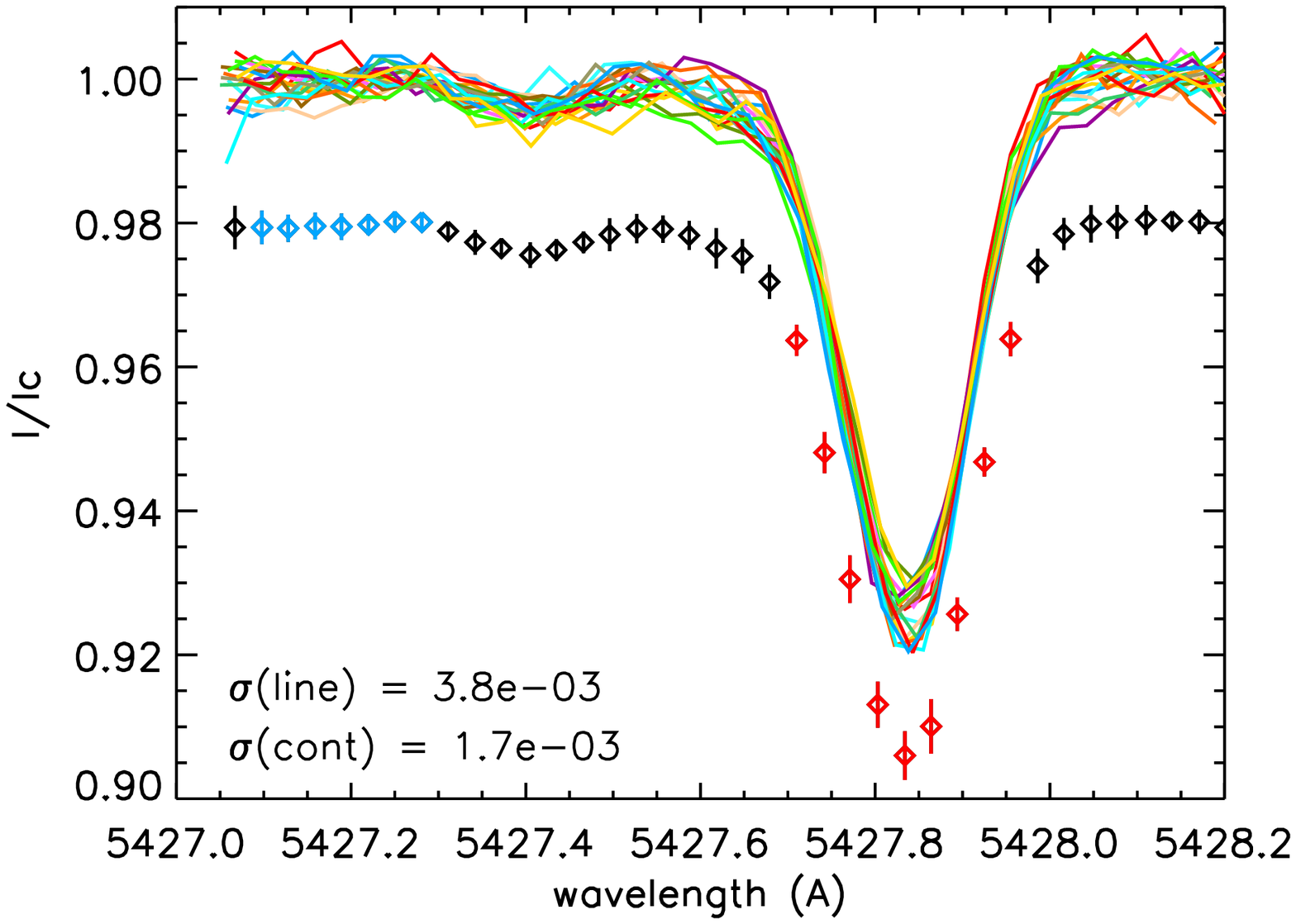}
\includegraphics[width=6cm]{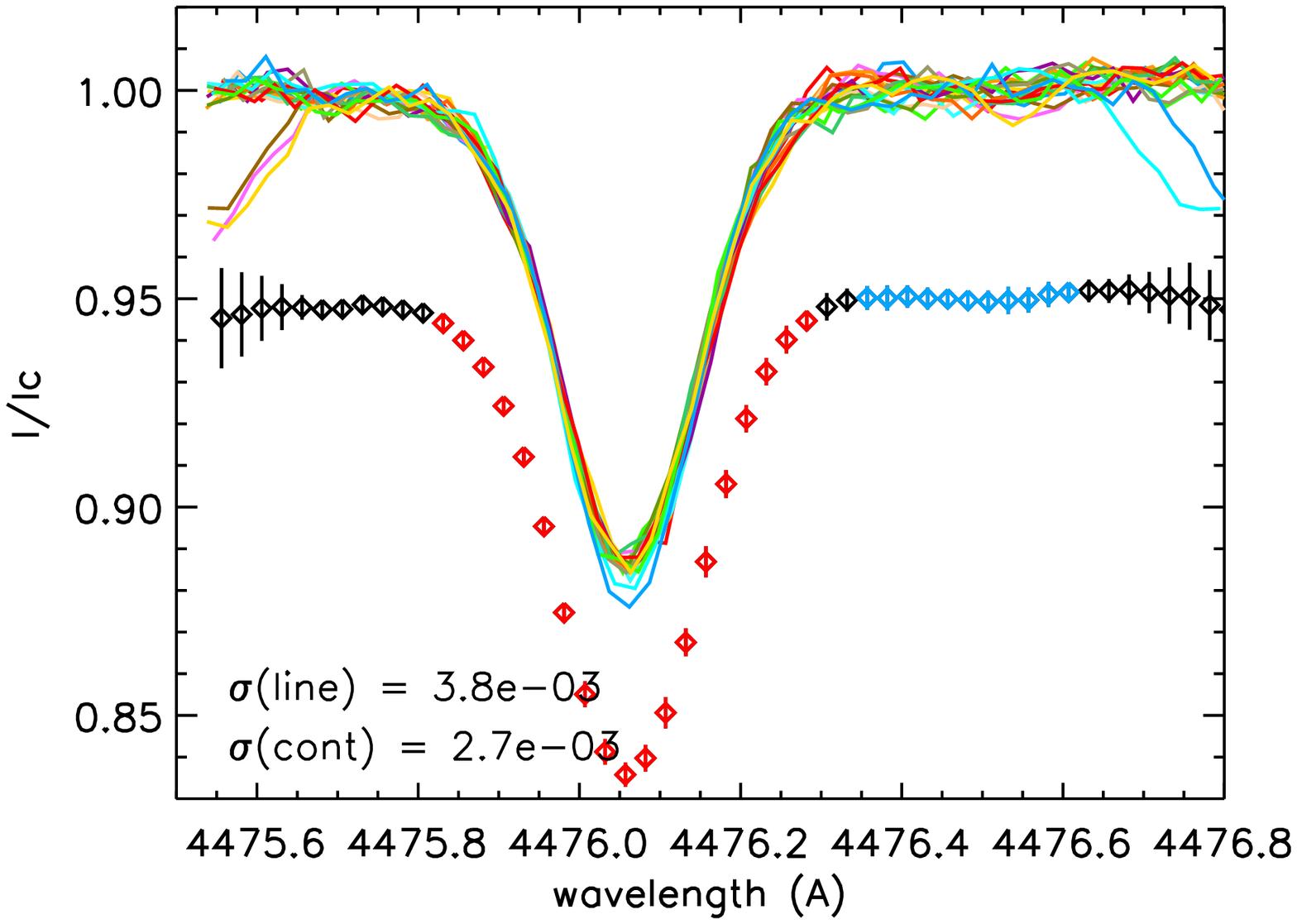}
\caption{\ion{Cr}{ii}\,5237\,\AA\ (left) and \ion{Fe}{ii}\,5427\,\AA\ (middle) lines of the primary, and \ion{Fe}{i}\,4476\,\AA\ (right) line of the secondary, for the 19 observations described in the text (Sec. 3.3.1, overplotted coloured lines). The mean (diamonds) and standard deviations (vertical bars) of each pixel inside and around the line have been overplotted and shifted by $-0.05$ in the y-direction for the left and right panels, and by $-0.02$ for the middle panelq. Red diamonds represent the pixels inside the spectral line, while blue diamonds represent the pixels in the local continuum. The averages of the standard deviations inside the line ($\sigma({\rm line})$) and in the continuum ($\sigma({\rm cont})$) are indicated in the bottom left corners.}
\label{fig:specvar}
\end{figure*}

\subsection{Extraction of the individual spectra}

For the following magnetic analysis, we need to extract individual spectra of both components separately for every observation. To this aim we first checked in the intensity spectra the variability of the lines of both components, and found that no strong variability is affecting our data. We can therefore efficiently remove the companion contamination from the combined spectra using the disentangled ones, producing individual spectra normalised to the summed continuum intensity of the binary. We finally normalised the spectra to the correct continuum intensity of the considered star (primary or secondary) using the fundamental parameters of Table \ref{tab:fitfp}. The whole procedure was first done using initial-estimate radial velocities, then done a second time using the radial velocities determined in Sec. 3.5. We describe below the steps of the procedure.

\begin{figure*}[t]
\centering
\includegraphics[width=7cm,angle=270,bb = 91 43 381 718,clip]{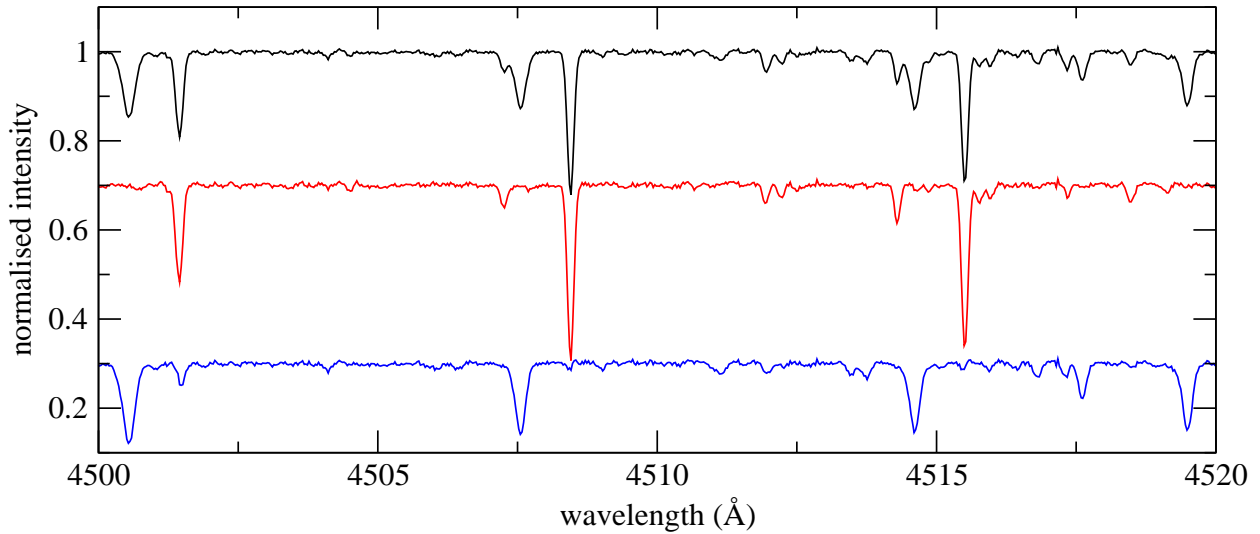}
\caption{Original combined (top, black) and re-normalised individual primary (middle, red) and secondary (bottom, blue) spectra of observation \#1. The primary and secondary spectra have been shifted downward for display purpose.}
\label{fig:subdis}
\end{figure*}

\subsubsection{Analysis of the variability of the spectra}

We first selected the 19 observations at orbital phases where the radial velocities between both components are larger than 30 \kms. This allows  us to study the variability inside the individual spectral lines of each component without contamination by spectral lines of the same transition from the companion. The temperature difference between both companions is large enough to also create blends of lines of different transitions even at large RV shifts, which is amplified by the abundance anomalies of both stars. We have therefore selected lines of the primary and secondary that are not blended with any lines from the companion whatever the orbital phase. The number of the selected lines is less than 10 per component and they correspond to Cr, Fe and Mn for the primary, and Fe and Ti for the secondary. For each line we proceeded as follows. We cancelled the orbital RV shift and superimposed the lines of all 19 observations. We finally measured the mean and standard deviation of each pixel inside the line, and compared them to the local continuum. An example of the overplotted spectral lines and the mean and standard deviation for a few transitions is illustrated in Fig. \ref{fig:specvar}. In the secondary's lines, the standard deviations of the line pixels are of the same order as that of the continuum, indicating no significant variation of the spectral lines over the 19 observations, while weak variability is observed in the lines of the primary. The expected main source of variability in Ap stars is of rotational origin. According to the ratio of radii ($R_{\rm A}/R_{\rm B}=1.24^{+0.04}_{-0.23}$) and the large uncertainties on the ratio of \vsini\ (\vsinia$/$\vsinib$=1.5^{+3.1}_{-0.9}$), there is a possibility that the system could be synchronised (i.e. $R_{\rm A}/R_{\rm B}=$\vsinia$/$\vsinib). We cannot therefore rule out the possibility that the orbital and rotation periods are identical. In that case, the variability of the lines cannot be tested for RV differences below 30\,\kms.

\subsubsection{Extraction and normalisation of the individual spectra}



The individual spectra of both components have been extracted in the same way for all observations. We have shifted the disentangled spectrum of the companion to the radial velocity of the companion for the relevant observation. We then subtracted the shifted disentangled spectrum from the observation. This provides the individual spectrum of the star normalised to the summed continuum of the binary ($I_{\rm cA}+I_{\rm cB}$). To correct it from the continuum intensity of the companion, we computed the flux ratio as follows. 
The difference in temperatures between the components is large enough for the flux ratio to be significantly dependent on wavelength. We used the fundamental parameters of Table \ref{tab:fitfp} and the ZEEMAN code to compute the flux ratio over the wavelength range of the disentangled spectra, as detailed in Sect. 3.2. 

Fig. \ref{fig:subdis} illustrates a portion of the spectrum of the binary (the original observation), and the extracted individual spectra of the primary and secondary. The extracted spectra do not show any signatures from the companion at the noise level. This is true for the whole wavelength range and all the observations, indicating the success of the applied procedure.

\subsection{Radial velocity determination}

\subsubsection{Least-squares deconvolution}

We applied the Least Squares Deconvolution (LSD) procedure to our spectra \citep{donati97}. This procedure combines the information contained in many metal lines of the spectrum, in order to extract the mean intensity (Stokes $I$) and polarised (Stokes $V$) line profiles. In Stokes $I$, each line is weighted according to its central depth, while in Stokes $V$ the profiles are weighted according to the product of the central depth, wavelength and Land\'e factor. These parameters are contained in a 'line mask' derived from a synthetic spectrum corresponding to the effective temperature and gravity of the star. The construction of the line mask for each star involved several steps. Our masks were first constructed using VALD line lists using stellar temperature, gravity, and abundances adapted to our stars. Our masks contain only lines with intrinsic depths larger or equal to 1\% of the continuum level. In a second step we excluded from the masks hydrogen Balmer lines, strong resonance lines, and lines contaminated with telluric lines. Finally, we modified the intrinsic line depths in order to take into account the relative depth of the lines of the observed individual spectra. The last two steps turned out to be crucial for a reliable determination of the radial velocities of both components, and therefore for ensuring the absence of residuals from the companion in the individual spectra after subtracting the companion spectrum. Finally, the mean Land\'e factor, the mean wavelength, and the mean line depth used to normalise the $I$ and $V$ profiles are respectively, 1.2, 500\,nm, and 0.2 of the continuum. The LSD S/N in $V$ are given in column \#2 of Table \ref{tab:bla} for the primary mask, and in column \#5 for the secondary mask.

\begin{table}[!ht]
\caption{LSD S/N and longitudinal field measurements}
\label{tab:bla}      
\centering          
\begin{tabular}{r|rr@{$\pm$}rr@{$\pm$}r|c}
\hline\hline
  & \multicolumn{5}{c|}{Primary} & Secondary \\
Sp. \# & S/N & \multicolumn{2}{c}{\bl ($V$)} & \multicolumn{2}{c|}{\bl ($N$)} & S/N\\
          &         & \multicolumn{2}{c}{G}             & \multicolumn{2}{c|}{G}            & \\
\hline
 1 & 11800 &  -17.8 & 4.2 &  4.4 & 4.2 & 26700 \\
 2 & 12200 & -20.6  & 4.8 &  2.8 & 4.8 & 27600 \\
 3 & 12500 & -26.4  & 4.2 &  0.3 & 4.2 & 28100 \\
 4 & 13800 &  -12.8 & 4.1 &  3.1 & 4.1 & 31000 \\
 5 & 14500 & -23.7  & 3.5 &  0.5 & 3.5 & 32800 \\
 6 & 15200 & -19.3  & 3.8 & -4.0 & 3.8 & 34400 \\
 7 & 14600 & -24.4  & 4.0 & -1.9 & 4.0 & 33000 \\
 8 & 14600 & -12.1  & 3.7 & -0.8 & 3.7 & 33100 \\
 9 & 15000 & -12.2  & 3.6 &  7.5 & 3.6 & 34100 \\
10 & 15000 & -20.9  & 3.8 &  5.4 & 3.8 & 33900 \\
11 & 13900 & -20.7  & 4.2 & -1.9 & 4.2 & 31400 \\
12 & 13400 & -13.8  & 4.3 &  0.8 & 4.3 & 30300 \\
13 & 14900 & -14.0  & 3.7 & -0.1 & 3.7 & 33900 \\
14 & 12200 & -14.9  & 4.4 &  2.4 & 4.4 & 27700 \\
15 & 14800 & -14.5  & 3.7 &  3.5 & 3.7 & 33400 \\
16 & 12700 & -23.7  & 4.4 &  0.8 & 4.4 & 28700 \\
17 & 13400 & -16.0  & 3.7 & -0.6 & 3.7 & 30400 \\
18 & 14300 & -18.7  & 4.0 & -4.6 & 4.0 & 32400 \\
19 & 14600 & -20.6  & 3.7 & -1.3 & 3.7 & 33100 \\
20 & 14500 & -15.8  & 3.9 & -1.8 & 3.9 & 32900 \\
21 & 10300 &  -19.0 & 5.6 & -7.4 & 5.6 & 23500 \\
22 & 14800 & -17.9  & 3.8 &  1.8 & 3.9 & 33500 \\
23 & 12400 & -16.4  & 4.4 &  5.5 & 4.4 & 28100 \\
24 & 15600 & -23.7  & 3.7 & -0.2 & 3.7 & 35200 \\
25 & 16000 & -24.3  & 3.7 & -0.1 & 3.7 & 36100 \\
\hline
\end{tabular}
\end{table}

\subsubsection{Fit of the LSD $I$ profiles}

We applied the LSD procedure to all our original observations of the binary system using successively the two masks generated for the two components. The temperature difference between both components is small enough that many spectral lines of the same transition are created by both components. As a result, the LSD $I$ profiles always display two components: a deep one corresponding to the mask used, and a shallow one corresponding to the companion. 

We performed a simultaneous Least-Squares fit to the 25 LSD $I$ profiles computed using the mask of the primary. Each profile is fit with the sum of two functions, each function modelling the line profile of one component. Each one of these two functions is the convolution of a rotation function (for which the projected rotational velocity \vsini\ is a free parameter in the fit) and a gaussian whose width is computed from the spectral resolution and an additional Gaussian broadening (free parameter) \citep{gray92}. 

\begin{figure}[!ht]
\centering
\includegraphics[width=4.3cm]{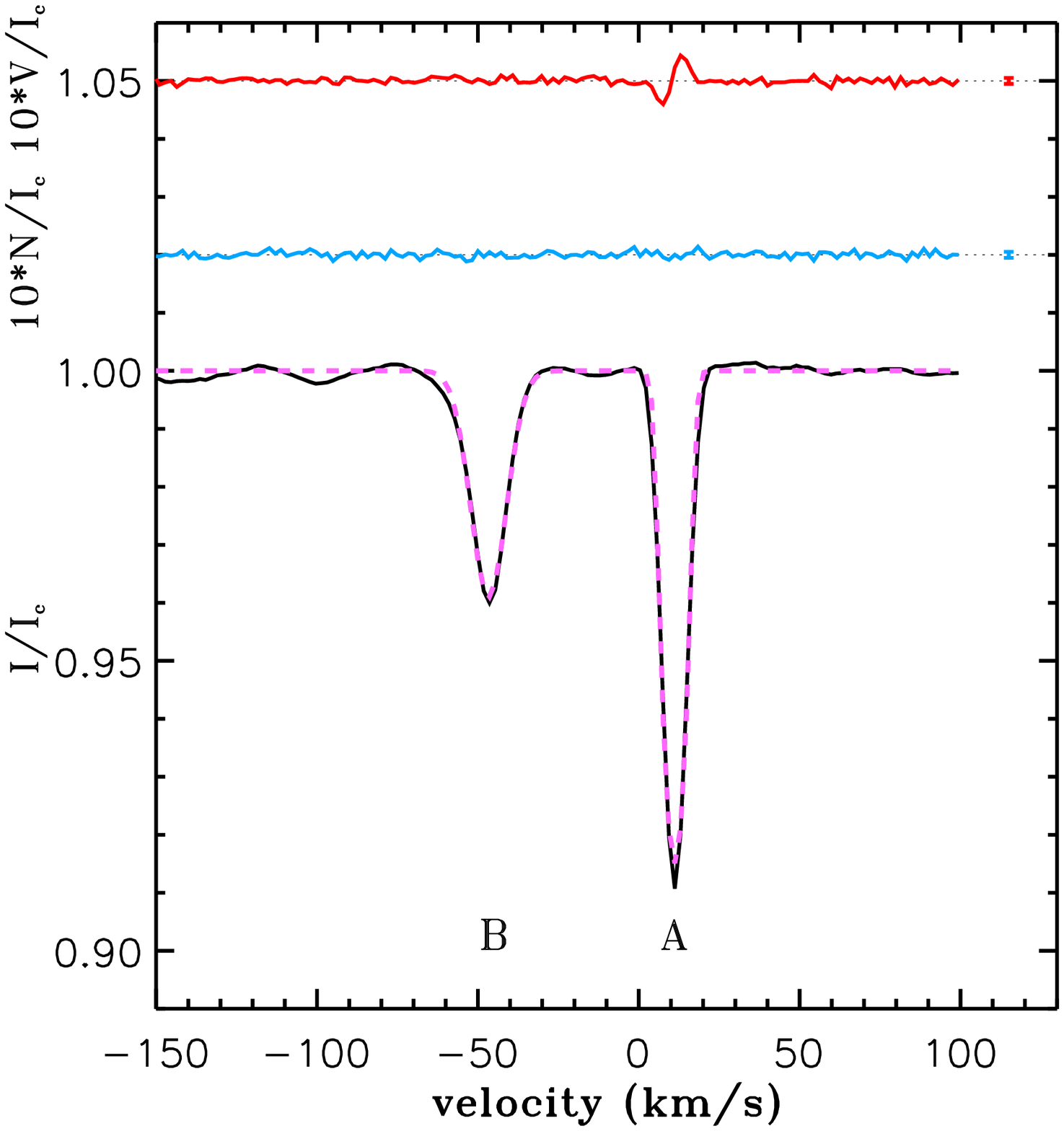}
\includegraphics[width=4.3cm]{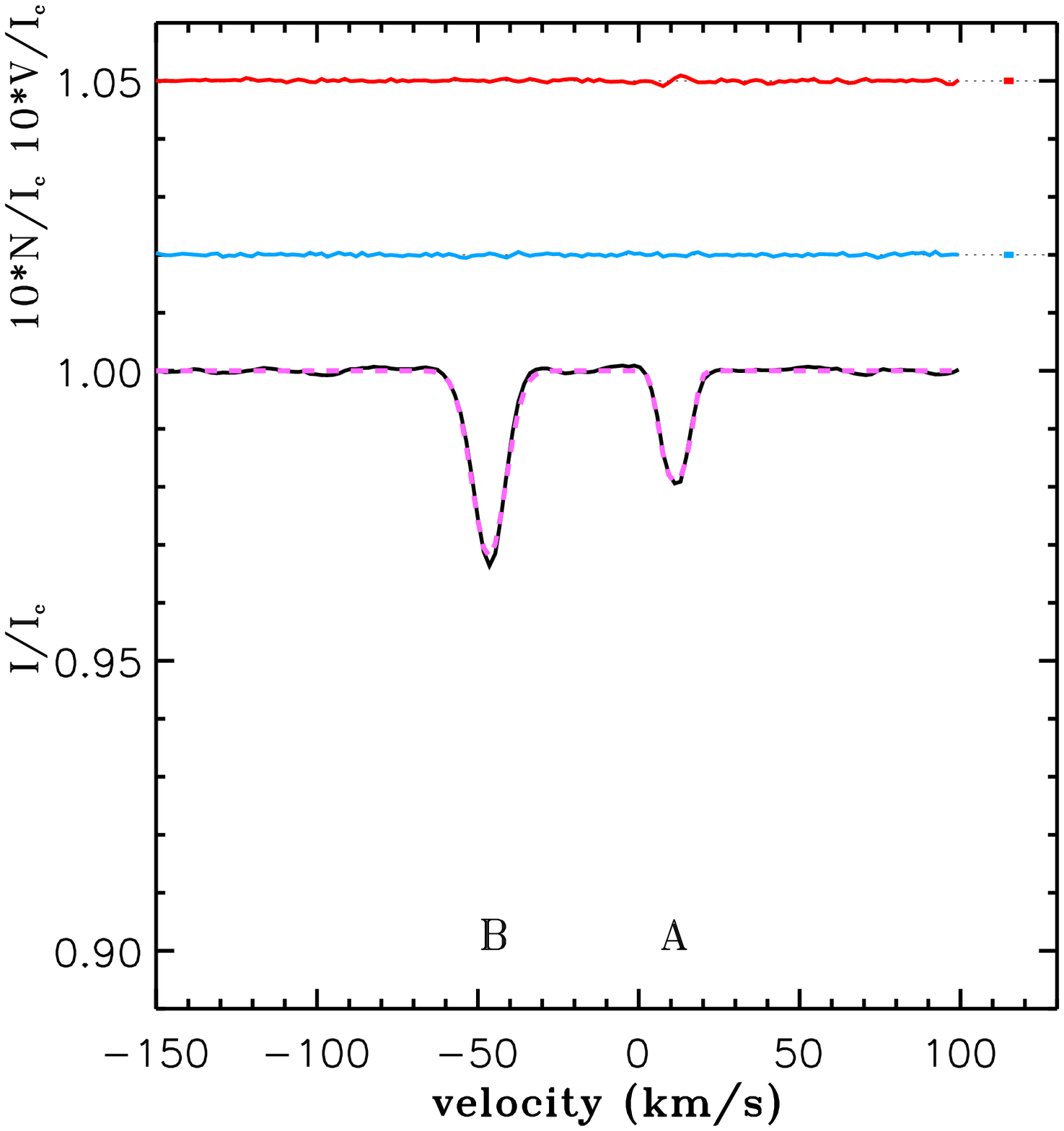}
\caption{LSD $I$ (bottom, black), $V$ (top, red), and $N$ (middle, blue) profiles of spectrum \#\,10 computed using the mask of the primary (left panel), and the mask of the secondary (right panel). The $V$ and $N$ profiles have been amplified and shifted in the y-axis for display purpose. The fits of the $I$ profiles (Sec. 3.4.2) are overplotted with dashed-pink lines.}
\label{fig:fiti}
\end{figure}

The free parameters of the fitting procedure are the centroids, depths, projected rotational velocities (\vsini), and the additional Gaussian broadening (\vmac) of both components. The centroids of both functions can vary from one profile to another, whereas the depths, \vsini\ and \vmac\ of both components cannot. This fitting procedure therefore assumes that the depths \vsini\ and \vmac\ of both components do not vary with time, which is reasonable when considering the faint variabilities observed only in a few individual lines of the primary. For both stars, the values of the additional Gaussian broadening are found to be non-zero, and are required to fit well the wings of the lines. The \vsini\ of both stars are indeed found to be relatively low ($<5$~\kms) and are clearly not dominating the broadening of the lines. As we have removed from the masks used to compute the LSD profiles all strong lines for which the pressure or natural broadening are dominent, we interpret this additional Gaussian broadening as a local thermal Doppler broadening. Such broadening is faint enough to be usually ignored in case of fast rotation but cannot be ignored in case of low rotation ($v\sin i \lesssim20$~\kms). Figure \ref{fig:fiti} shows an example of such a fit for spectrum \#10, for both masks. This automatic fitting procedure enables us to measure the radial velocities with 1$\sigma$ errors of $1.6-1.8$\,\kms for the primary (using the primary mask), and $2-3$\,\kms for the secondary (using the secondary mask). Table \ref{tab:vrad} summarises our adopted radial velocity measurements for both components, while the \vsini\ and \vmac\ are summarised in Table \ref{tab:fitfp}.

\begin{table}[!ht]
\caption{Radial velocity measurements}
\label{tab:vrad}      
\centering          
\begin{tabular}{rr@{$\pm$}rr@{$\pm$}r}
\hline\hline       
Sp. \# & \multicolumn{2}{c}{\vrad$_{\rm A}$} & \multicolumn{2}{c}{\vrad$_{\rm B}$} \\
          & \multicolumn{2}{c}{\kms}                    & \multicolumn{2}{c}{\kms}                \\
\hline
 1 &   10.6 & 1.8 & -49.0 & 2.4 \\
 2 &  -30.4 & 1.8 &  14.2 & 3.2 \\
 3 &  -36.1 & 1.7 &  23.3 & 3.1 \\
 4 &   11.5 & 1.7 & -50.6 & 3.0 \\
 5 &   -3.7 & 1.8 & -27.2 & 3.1 \\
 6 &  -26.2 & 1.7 &   8.2 & 2.9 \\
 7 &  -36.9 & 1.8 &  26.1 & 3.1 \\
 8 &   -7.6 & 1.7 & -20.2 & 2.4 \\
 9 &    9.9 & 1.8 & -48.9 & 2.3 \\
10 &   10.3 & 1.6 & -47.4 & 2.3 \\
11 &  -29.7 & 1.8 &  13.7 & 2.5 \\
12 &  -37.3 & 1.8 &  25.0 & 2.4 \\
13 &  -26.3 & 1.7 &   8.7 & 2.6 \\
14 &   -3.6 & 1.8 & -27.3 & 2.6 \\
15 &   11.4 & 1.7 & -50.8 & 2.4 \\
16 &  -20.9 & 1.8 &  -0.4 & 2.3 \\
17 &   12.6 & 1.7 & -51.3 & 3.1 \\
18 &  -33.3 & 1.8 &  19.2 & 2.4 \\
19 &    7.6 & 1.7 & -41.8 & 2.8 \\
20 &   12.8 & 1.7 & -50.7 & 2.1 \\
21 &  -22.7 & 1.8 &   2.9 & 2.5 \\
22 &  -31.4 & 1.7 &  17.3 & 2.6 \\
23 &    7.9 & 1.7 & -45.2 & 2.7 \\
24 &   -5.2 & 1.7 & -24.0 & 2.3 \\
25 &  -26.1 & 1.7 &   8.3 & 2.5 \\
\hline
\end{tabular}
\end{table}

\begin{figure}[!ht]
\centering
\includegraphics[width=9cm]{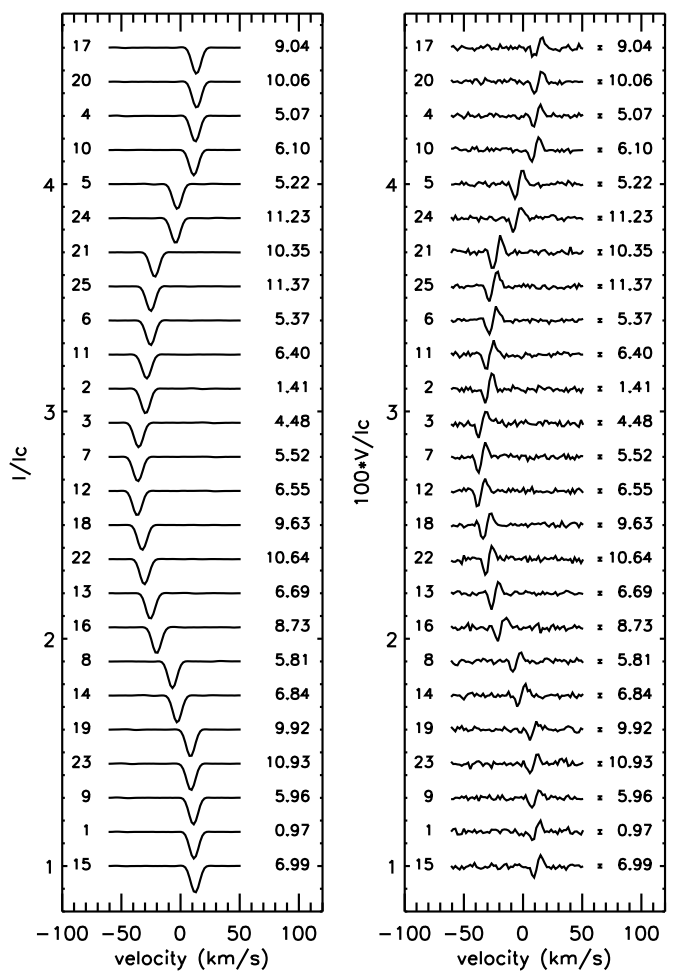}
\caption{LSD $I$ (left) and $V$ (right) profiles of the primary ordered by orbital phase. The $V$ profiles have been amplified for display purpose. The spectrum number, mean errors, orbital cycles and phases are indicated on the left and right side of the profiles. The orbital cycles have been subtracted by 805, the cycle of the first observation of our dataset.}
\label{fig:alllsd}
\end{figure}

\section{Magnetic analysis of the primary star}

\subsection{The LSD $V$ and $N$ profiles}

The LSD $V$ profiles of our observations (see an example in Fig. \ref{fig:fiti}) show clear signatures in the primary component, but not in the secondary. The false-alarm probabilities of the primary signatures, as defined by \citet{donati97}, are all lower than $10^{-9}$ indicating that the signatures are all significant (definite detections). The signatures display a typical Zeeman S-shape. They are centred on the centroid of the primary component, and are as broad as the primary $I$ profiles. The $N$ profiles display flat signal in all our observations indicating that the $V$ profiles are not contaminated with spurious signal. All of which allows us to confidently conclude that a magnetic field is detected at the surface of the primary, but not at the surface of the secondary. If a magnetic field is present at the surface of the secondary, it creates a signature that is lower than the noise level of our observations. We can therefore assume that the contribution of the secondary to the circular polarisation signal is negligible, and that the $V$ profiles are dominated by the signal from the primary. However, the $V$ profiles do need to be corrected for the continuum of the secondary as they are computed with respect to the summed continuum intensity of the binary. We performed this re-normalisation in the same way as for the intensity spectra (Sec. 3.3.2) for the $V$ and $N$ spectra.

To analyse the magnetic field of the primary we used the LSD profiles computed with the mask of the primary, the individual spectra of the primary (as extracted and normalised in Sec. 3.2.2), and the re-normalised $V$ and $N$ spectra described above. Fig. \ref{fig:alllsd} shows the $I$ and $V$ profiles of all our observations ordered by orbital phase. The $V$ profiles all display a similar S-shape. The amplitudes change slightly, but the sign is always the same. This indicates that only the magnetic South pole was visible at the surface of the star at the time of the observations.

\subsection{Determination of the rotation period}

The magnetic fields of Ap stars are in most cases dominated by a strong dipolar component. In a few cases a quadrupolar or octupolar component can be as large or even larger than the dipolar one, but in any case the dominant component is a low-order multipolar field. The magnetic axes are usually inclined with respect to the rotation axis, causing rotational modulation of the longitudinal magnetic field at the surface. We can therefore expect periodic variability in our data directly linked to the rotation period at the surface of the star. We can use this variability of the $V$ profiles to find the rotation period. To this aim we applied various methods described below.

\subsubsection{Search for the rotation period using the longitudinal field measurements}

We measured the line-of-sight component of the magnetic field integrated over the stellar surface (the longitudinal field \bl) from our data by computing the first order moment of $V$ and the equivalent width of $I$ using the method described by \citet{wade00}. We used integration limits of 23\,\kms\ wide centred on the radial velocity of the $I$ profiles. The values we find are all negative, and range from $\sim$$-26$\,G to $\sim$$-12$\,G, with typical error bars of about 4\,G. Using the same method, we measured the longitudinal field values in the LSD $N$ profiles. All values are found to be consistent with 0~G, confirming the absence of spurious polarisation in our data. The \bl\ measurements for $V$ and the pseudo-\bl\ measurements for $N$ are given in columns \#3 and \#4 of Table \ref{tab:bla}.


We searched for periodic variations in the $B_l$ values using two methods: a simple sine fit to the $B_l$ values, and a clean method. The software we used for these searches are described in \cite{gutierrezsoto09}. The sine fit provided two significant periods at 5.80 and 6.82 d. The clean method provided peaks around 5.68, 6.45, 2.63, 3.55\,d. None of these periods are identical, we conclude that no periodicity can be found in the $B_l$ values. However, we note that one of the periods we found (6.82 d) is close to the orbital period.

\subsubsection{Search for the rotation period using the Stokes $V$ profiles}

Finally, we attempted a search for periodicity using the entire $V$ profiles, using both model-independent and model-dependent methods.

\vspace{30pt}
{\bf \noindent Model-independent methods}

We first performed a frequency analysis using various methods. We first used the FAMIAS package\footnote{The software package FAMIAS was developed in the framework of the FP6 European Coordination Action HELAS (http://www.helas-eu.org/)} \citep{zima08}. Using a mean 1D Fourier technique applied to the LSD Stokes $V$ profiles and a S/N significance criterion of 4, we obtained a main frequency peak at $f_1$ =1.14533 c.d$^{-1}$ with S/N=18, and its 1-d aliases at $\sim$0.14, 2.14, 3.14, etc. We also detect peaks at 1, 2, 3,... c.d$^{-1}$ and combinations of these peaks with $f_1$ at 0.856 c.d$^{-1}$, etc. Then, we applied the mean 1D Fourier method of FAMIAS to the LSD $I$ profiles. We found a frequency at 0.14745 c.d$^{-1}$ with S/N=9. Again we also find 1-d aliases and combinations. From this frequency analysis of the $V$ and $I$ profiles, we conclude that $f\sim$[0.145-0.148] c.d$^{-1}$ is the main frequency of variations of HD~5550, corresponding to a period comprised between 6.75 and 6.90\,d.

\begin{figure}[!ht]
\centering
\includegraphics[width=9cm]{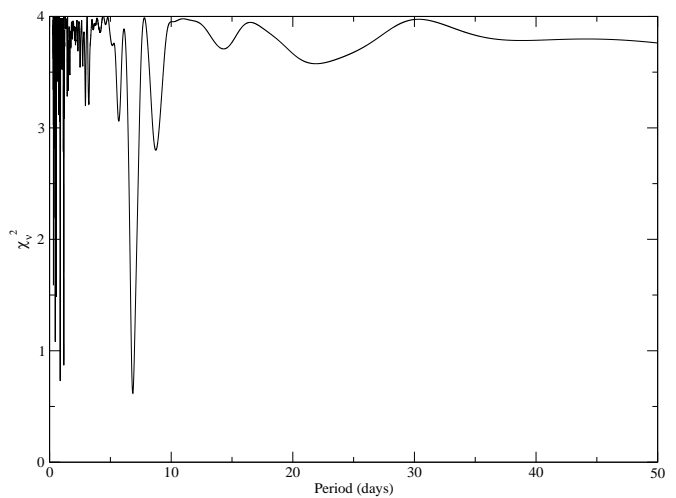}
\caption{Periodogram of the LSD $V$ profiles of the magnetic Ap component.}
\label{fig:periodo}
\end{figure}

We then fit the individual variation of each pixel within the $V$ profiles (within $\pm 10$~\kms of the line center) with a sinusoidal function with three free parameters (amplitude, zero point and phase) with a fixed period. This was done for each pixel, for many periods between between 0.3 and 50 days, constructing a periodogram in $\chi^2$ and period for that pixel. We then computed the average of all the periodograms, and looked for mimima in $\chi^2$ (Fig. \ref{fig:periodo}). 
Only one significant minimum at a 5-$\sigma$ confidence level is found above 1.2~d, at a period of $6.84\pm0.15$\,d with a reduced $\chi^2$ of 0.62. Below 1.2~d there are significant aliasing. However, those periods are highly improbable with the observed \vsini, and the magnetic and binary nature of the target. Furthermore, the temporal sampling of our observations suggests we do not have much sensitivity below 1.2~d.

\vspace{11pt}
{\bf \noindent Model-dependent method}

\begin{figure}[!t]
\centering
\includegraphics[width=11cm,bb = 100 150 700 600,clip]{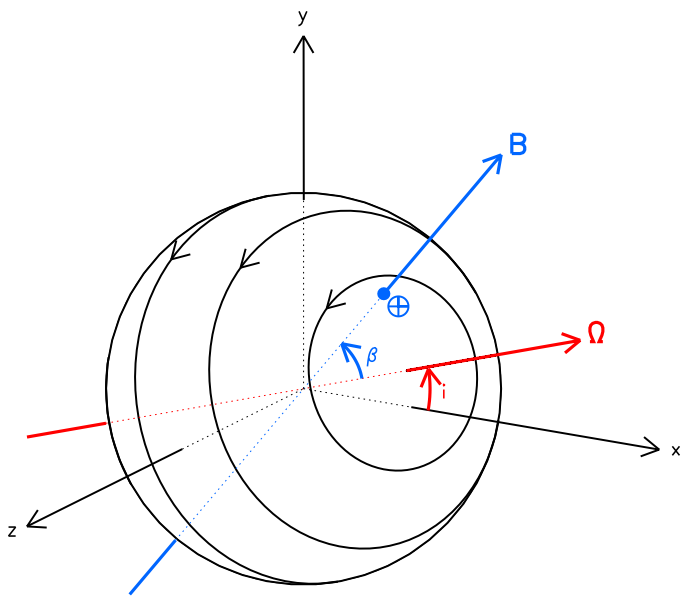}
\caption{A 3D schematic view of the magnetic star at the rotation phase 0.0 of our modeling (see the text). The rotation (red, labelled {\bf $\Omega$}), magnetic (blue, labelled {\bf B}), and cartesian ($x$,$y$,$z$) axes are represented outside (full lines) and inside the star (dotted lines). The $x$ axis is pointing towards the observer. The arrows representing the inclination angle ($i$) and magnetic obliquity ($\beta$) angles point towards positive values. The parallels at 30, 60 and 90\ddeg\ of latitude are plotted with respect to the rotation axis for display purpose. The arrows superimposed with the parallels indicate the direction of rotation of the star. The positive magnetic pole ($\oplus$) is plotted as a large blue dot at the intersection of the magnetic axis and the surface of the star.}
\label{fig:model}
\end{figure}

The last method we employed has already been described in detail in \citet{alecian08}. We assume that the magnetic field of the primary component is a dipole, centered inside the star, and inclined to the rotation axis by an obliquity angle $\beta$. The rotation axis is inclined with respect to the line-of-sight by an angle $i$, and the magnetic strength at the pole is defined as \bd\ (Fig. \ref{fig:model}). The magnetic pole points towards the observer at phase 0.0 (given by the initial epoch $T_0$). If both positive and negative poles are observed over one rotation cycle, then we choose the positive magnetic pole to point towards the observer at phase 0.0. We compute the local intensity line profiles on each point of a 5041 elements grid simulating the observed surface of the star. The local $I$ profile is a Gaussian with a width dependent on the instrumental and local thermal broadening, shifted to the local radial velocity according to the given \vsini. The local $V$ profiles are computed in the weak field approximation using the local $I$ profiles, a mean wavelength of 500 nm, and a mean Land\'e factor of 1.2, equal to the normalised values used in the LSD computation. The $V$ and $I$ profiles are then integrated over the surface using a linear limb-darkening law with a coefficient of 0.43, suitable for the temperature of the star \citep{claret11}, and for given values of the rotation phase ($\phi$), $i$, $\beta$ and \bd. The rotation phase is obtained from the rotation period (\prot), $T_0$, and the HJD of our observations. For all our observations, we computed a grid of Stokes $V$ profiles for various values of \prot\ (between 1 and 66 d), $\phi_0$ (the phase shift between $T_0$ and the HJD of our first observation, varying from 0 to 1), $i$ (between 0\ddeg\ and 180\ddeg), $\beta$ (between 0\ddeg\ and 180\ddeg), and \bd\ (between 20 and 1000 G). We then computed the $\chi^2$ on each point of the grid. Next, we searched for the minimum of $\chi^2$ as a function of the rotation period. We find a periodogram very similar to Fig. \ref{fig:periodo}, with only one significant minimum at 6.84\,d, with a reduced $\chi^2$ of 1.3, and aliasing below 1.2\,d.

\subsubsection{Conclusions from the rotation period search}

We used many different methods to attempt to find a period in our $V$ profiles, as well as in our $I$ profiles (assuming that the faint variabilities observed in the $I$ profiles are of rotational origin). We found no common period to all the methods. On the other hand, all the methods we employed using the entire Stokes $V$ profiles provide us with a common period at 6.84\,d. Assuming this period, the evolution of the profiles as a function of phase appears coherent with a simple dipolar field for all the profiles.


We estimate the uncertainties of the rotation period at a 3$\sigma$ confidence level to be $6.84_{-0.39}^{+0.61}$\,d. This indicates that the rotation period is similar to the orbital one ($P_{\rm orb}=6.82054$\,d), which would imply that the system is synchronised. We have already reported above that our measurements of the \vsini\ and the ratio of radii permits the system to be synchronised. Furthermore, the system is almost circularised ($e=0.005$), and the evolutionary models of a binary system predict that the synchronisation time is shorter than the circularisation time \citep{zahn77}. It is therefore very likely that HD 5550 is already synchronised, supporting our rotation period determination.

\begin{figure*}[!ht]
\centering
\includegraphics[width=5cm,angle=90]{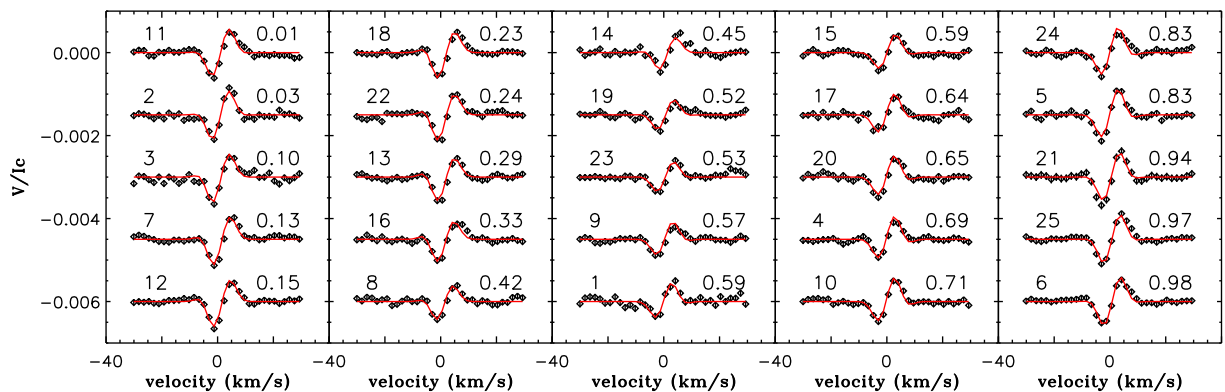}
\caption{Observed (diamonds) and modelled (red lines) LSD $V$ profiles of the primary ordered by rotational phase. The spectrum number and  the rotational phases are indicated on the left and right side of the profiles, respectively.}
\label{fig:alllsd}
\end{figure*}

\subsection{The magnetic topology of the primary}

We assume a rotation period of 6.84\,d, and propose a solution for the magnetic topology of the primary. The oblique rotator model used to fit the $V$ profiles has already been described in Sect. 4.2.3. The best model corresponds to $i=32_{-38}^{+21}$\,\ddeg, $\beta=156\pm17$\,\ddeg, and $B_{\rm d}=65 \pm 20$ G, where the error bars correspond to a 3$\sigma$ confidence level, based on $\chi^2$ statistics. The rotational initial epoch is found to be $T_0=2\,456\,481.5 \pm 0.8$, which corresponds to a phase shift from the orbital initial epoch of about 0.6. In order to check the uniqueness of the solution we plotted $\chi^2$ maps as a function of two of the four fitted parameters ($T_0$, $i$, $\beta$, \bd), by fixing the other parameters to the values at minimum. We plotted a total of 6 maps corresponding to all possible combinations of 2 parameters among 4, and found no other $\chi^2$ minimum. The $V$ profiles corresponding to this solution are plotted over the observations in Fig. \ref{fig:alllsd}. We find that the model reproduces well the observations. The uncertainties on the parameters are rather large, which is a consequence on the faint variability of the $V$ profiles over the rotation cycle. 

{ The uncertainty on the flux ratio between both components computed in Sec. 3.3 has not been taken into account in the above study. The uncertainty on the flux ratio is controlled by the uncertainty on the radii ratio ($R_{\rm A}/R_{\rm B}$) and on the temperatures ratio ($T_{\rm effA}/T_{\rm effB}$). The former is the dominant uncertainty, and does not depend on wavelength, while the later is well constrained, but is dependent on the wavelength. As both $I$ and $V$ profiles are computed with respect to the continuum, the uncertainty on the radii ratio has no impact on the magnetic model we derived (the continuum terms effectively cancel out). However, the wavelength dependency of the temperature ratio can affect the solutions of the LSD calculations. To estimate the impact of the uncertainty of the temperature ratio on our solution we have computed $I$ and $V$ LSD profiles of the primary without correcting the continuum for the secondary's contribution, simulating a very extreme case with a temperature ratio of 1. We find a solution very similar to the above solution, and consistent within the error bars. We can therefore neglect the uncertainty on the temperature ratio in our final solution.}


\section{Upper limit on the magnetic field of the secondary star}


As detailed above, we do not detect a magnetic field at the surface of the secondary component of HD\,5550 in our data. The longitudinal field values we measured by integrating the LSD $I$ and $V$ profiles around $\pm12$\,\kms of the line center, are all consistent with zero with uncertainties comprised between 3 and 4\,G.

The signature of a weak magnetic field may have remained hidden in the noise of the spectra of the secondary component. We propose to estimate its maximum
strength assuming the simplest magnetic configuration, i.e. a dipole. To this aim, we need to isolate LSD $I$ and $V$ profiles of the secondary without any contamination from the primary for as many observations as possible. Unfortunately, as described above, the LSD profiles, computed with the line-mask tailored for the secondary, are contaminated by the primary. This is problematic near conjunction when both line components are partially or entirely superimposed. While it is possible to disentangle the signals of both components in the intensity spectrum because both components display significant signal, when one of the two components displays no signal (as in our $V$ spectra) it is not possible to perform a similar disentangling. Therefore, we only used the 15 spectra for which the secondary component is well separated from the primary (i.e. profiles \#1, 2, 3, 5, 6, 9, 10, 11, 12, 15, 17, 19, 21, 22 and 25), so that the secondary's LSD $V$ profiles are not contaminated by the magnetic field of the primary. 

To determine the upper limit on the magnetic field of the secondary star, we first fitted the LSD $I$ profiles using a model computed as follows. We first computed local $I$ profiles with two superimposed Gaussian centred on the same velocity, then integrated them over the observed surface. The free parameters of the fitting procedure are the broadenings and depths of the two Gaussians, the radial velocity and \vsini. The two Gaussians are required in order to reproduce both the wings and the core of the LSD $I$ profiles. The derived \vsini\ are similar to the one derived in Sec. 3.4.2. We then computed 1000 synthetic $V$ profiles for various values of the polar magnetic fields \bd\ in the way described in Sec. 4.2.3, using the same mean Land\'e factor (1.2) and wavelength (500 nm) as the observations. Each of these models uses a random inclination angle $i$, obliquity angle $\beta$, and rotational phase. We then added a white Gaussian noise to each profile with a null average and a variance corresponding to the S/N of the observed profile. We proceeded this way for the 15 observations, providing us with 15 sets of 1000 models of $V$ profiles.

\begin{figure}[!ht]
\begin{center}
\resizebox{\hsize}{!}{\includegraphics[clip,angle=-90]{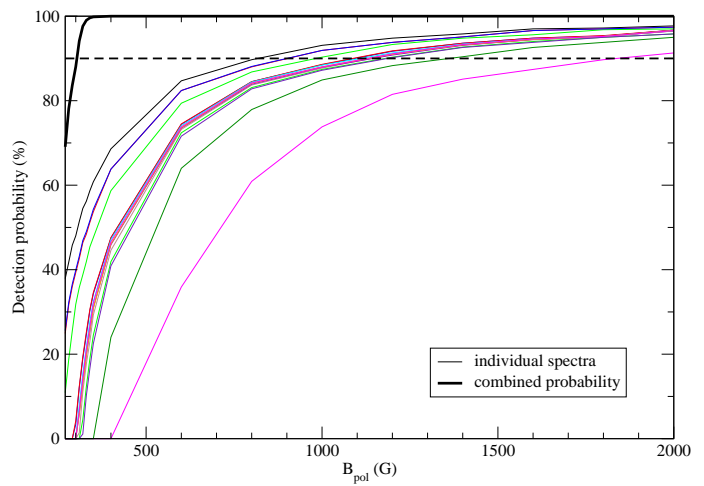}}
\caption[]{Detection probability of a magnetic field in each spectrum of the
secondary component of HD\,5550, as a function of the magnetic polar field
strength. The horizontal dashed line indicates the 90\% detection probability, and the thick curve shows the combined probability.}
\label{limit}
\end{center}
\end{figure}

For each set of 1000 models, we computed the probability of detection of a magnetic field in each simulated $V$ profile by applying the Neyman-Pearson likelihood ratio test \citep[see
e.g.][]{helstrom1995,kay1998,levy2008}. This allows us to decide between two
hypotheses: the profile contains noise alone or it contains a noisy Stokes $V$
signal. This rule selects the hypothesis that maximises the detection
probability, while ensuring that the false-alarm probability $P_{\rm FA}$ is not
higher than a prescribed value. We used $P_{\rm FA} = 10^{-3}$ for a marginal
magnetic detection, as commonly assumed in the literature
\citep[e.g.][]{donati97}. We then calculated the number of detections among the
1000 models for each one of the 15 observations and for each value of the field strength, and plotted the detection rates (number of detections over 1000) as a function of field strength (see Fig.~\ref{limit}).

We require a 90\% detection rate to consider that the field would statistically
be detected in one observation. This allows to determine the upper limit of the field strength below which we are not able to detect it in our observations. The upper limits we find for each profile vary between 130 and 200\,G, depending on the quality of our data.

Since 15 spectra are at our disposal, statistics can be combined to extract a stricter upper limit by computing the probability that a magnetic field of a given strength has been detected in at least one of the observations. If we assume that the observations are all independent, we can compute this combined probability following the method of \citet{neiner2015}, that we remind here. The computation of the detection rate performed for each observation can be considered as a random test $i$, with $0<i<n$, and $n=15$ is the number of observations. Each test can have only two outcomes: detected (D) or not-detected (ND) with probabilities $P_i(\rm D)$ and $P_i(\rm ND)$, where $P_i(\rm D)$ is  the detection rate described above and $P_i(\rm ND) = 1-P_i(\rm D)$. If $n$ tests are performed, all independently, among all the combinations of outcomes, there is only one where all individual outcomes are ND with a probability $\prod_{i=1}^{n}P_i(\rm ND)$. All other combinations will contain at least one D. The probability to have at least one detection in a set of $n$ independent observations is $1-\prod_{i=1}^{n}P_i(\rm ND)=1-\prod_{i=1}^{n}(1-P_i(\rm D))$. We have computed this combined probability for many different polar field strengths (thick curve in Fig. \ref{limit}) and applied again the 90\% limit to derive the upper limit. The final upper limit above which a dipolar magnetic field at the surface of the secondary component would have been detected in our data is 40\,G.

\section{Summary and discussion}

In the context of the BinaMIcS project, we have acquired high-resolution spectropolarimetric Narval observations of a binary system composed of two A-type stars. We have analysed the intensity and circularly polarised spectra of both components, and derived the fundamental parameters of both components. We find in the primary that the abundances of iron-peak elements (except Cr) are slightly enhanced compared to the solar ones \citep{asplund09}, that Cr is clearly overabundant, and that rare-earth elements are strongly overabundant, typical of Ap peculiarities. This confirms the Ap nature of the primary suspected by \citet{carrier02}. In the secondary, overabundance of the iron-peak elements, extreme overabundance of Ba, and underabundance of the Ca demonstrate its Am nature. This classification, based only on their chemical properties, is consistent with the magnetic detection in the primary only. Indeed, it is now well accepted that only strong magnetic fields (with field strengths higher than $\sim$300\,G) are present only in the Ap/Bp stars, while the Am stars have no  or very weak magnetic fields \citep[e.g.][]{shorlin02,auriere07,blazere15}.

The magnetic field topology of the primary is typical of Ap/Bp stars, i.e. a configuration dominated by a dipolar component with a non-null magnetic obliquity between the rotation and magnetic axes. The dipolar field strength (65~G) is however the lowest ever found in an Ap star. Indeed, typical magnetic strengths of the Ap and late-type Bp stars are of the order of 1~kG, with a range between 300~G and 30~kG \citep[e.g.][]{borra80,landstreet92,bagnulo06}. The latest studies of the magnetic fields in A-type stars provided strong evidence for a magnetic dichotomy between the strongly magnetised Ap stars (with fields larger than $\sim$100\,G), and the ultra-week magnetic Vega-like stars (with fields lower than 1~G), with a clear desert between both magnetic strength limits \citep[e.g.][]{auriere07,auriere10,lignieres09,petit11}. On the other hand, the Ap component of HD~5550 has a higher temperature than the samples analysed in those works. In early-B type stars (above B3) polar field strengths of 10-100\,G have been recently found \citep[Briquet, Neiner, Petit et al. submitted]{fossati15}. While the limit of 100\,G appears to be strict at temperatures lower than 10000\,K, it may decrease with increasing temperatures. In any case,
the magnetic field of the Ap component of HD 5550 is still more than an order of magnitude larger than the ultra-week magnetic A stars, and do not challenge the dichotomy previously established. The upper limit on the magnetic field of the secondary that we derived is 40~G. This is compatible with the ultra-weak fields detected in some Am stars, and is also consistent with the magnetic dichotomy of the A/B stars.

This is only the second time this kind of magnetic analysis has been performed on a short-period double-lined spectroscopic binary containing at least one magnetic Ap component. The other system is HD 98088 \citep{folsom13} with a magnetic Ap primary that is cooler ($T_{\rm eff}=8300$~K) than in HD~5550, and a non-magnetic Am secondary with a temperature similar to HD\,5550\,B ($T_{\rm eff}=7500$~K). The orbital periods of the systems are rather close (5.9\,d vs 6.8\,d). Both components of HD 98088 are already tidally locked. Similarly, the primary of HD 5550 is likely already synchronised with the orbit. \citet{folsom13} found that the Ap component hosts a predominantly dipolar field with a polar strength of about 3.8 kG, and a magnetic axis pointing approximately towards the companion. In the case of HD 5550, we do not have any observational constraint on the orbital axis, but if the primary is synchronised it is likely that both rotational and orbital axis are aligned. Within this assumption, as the South pole of the magnetic axis is pointing towards the observer at an orbital phase of 0.6, the magnetic axis and the line joining both components are not in the same plane. Furthermore, the angle between the rotation and magnetic axis being about 24\ddeg, even if the errors are large, the magnetic axis is clearly not aligned with the orbital plane. Therefore, along the orbit, the secondary sees the magnetic axis under the same and large angle, and only the positive pole.

It is interesting to note that the weakest magnetic Ap star is a member of a close binary system. While this result could be a pure coincidence, it may also indicate that fossil magnetic fields and star formation processes are linked. Indeed, close binary systems must have been formed during the star formation process. However, theoretical works have shown that in presence of strong magnetic fields, the fragmentation is reduced, and binary systems are difficult to form \citep[e.g.][]{commercon11}. Therefore, binary systems can only be formed in relatively weakly magnetised clouds, while only single stars can form in relatively strongly magnetized clouds. If a memory of the initial conditions of star formation are retained all along the star formation processes and star lifetime, only relatively weak magnetic fields can be found in close binary systems.
It is too early to be firm about this interpretation. Once the BinaMIcS collaboration has accumulated a larger number of analyses of this type, we will be able to draw a larger picture on how magnetism can impact the binary evolution, and conversely.

\begin{acknowledgements}
We thank the referee for his judicious comments. We acknowledge financial support from " le Programme National de Physique Stellaire" (PNPS) of CNRS/INSU, France. EA acknowledge financial support from "le Laboratoire d'Excellence OSUG@2020 (ANR10 LABX56)" funded by "le programme d'Investissements d'Avenir".
\end{acknowledgements}


\bibliographystyle{aa}
\bibliography{hd5550}

\end{document}